\documentclass {elsart}

\usepackage{graphics}
\usepackage{graphicx}
\usepackage{epsfig}

\usepackage{amssymb}

\begin{document}

\begin{frontmatter}



\title{Encryption of Covert Information into Multiple Statistical
Distributions}


\author{R. C. Venkatesan}

\address {Systems Research Corporation \\
I.T.I Rd., Aundh, Pune 411007, India \\
Email: ravi@systemsresearchcorp.com }\
\begin{abstract}
A novel strategy to encrypt covert information (code) via unitary
projections into the \textit{null spaces} of ill-conditioned
eigenstructures of multiple host statistical distributions, inferred
from incomplete constraints, is presented. The host \textit{pdf}'s
are inferred using the maximum entropy principle.  The projection of
the covert information is dependent upon the \textit{pdf}'s of the
host statistical distributions.  The security of the
encryption/decryption strategy is based on the extreme instability
of the encoding process.  A self-consistent procedure to derive keys
for both \textit{symmetric} and \textit{asymmetric} cryptography is
presented. The advantages of using a multiple \textit{pdf} model to
achieve encryption of covert information are briefly highlighted.
Numerical simulations exemplify the efficacy of the model.
\end{abstract}

\end{frontmatter}

\section{Introduction}
The success of prominent contemporary cryptosystems is attributed to
the degree of difficulty in computing integer factorizations [1] and
discrete logarithms [2, 3], respectively. This paper describes a
novel strategy to encode covert information via unitary projections
into the \textit{null spaces} of the eigenstructures of a hierarchy
of host statistical distributions (multiple \textit{pdf}'s).  The
multiple \textit{pdf}'s are inferred from an incomplete set of
constraints (physical observable's) using the maximum entropy
(MaxEnt) principle [4]. Here, a multiple \textit{pdf} model is
defined as a hierarchial ensemble of \textit{pdf}'s $
p^\nu;\nu=1,... $.  The index $ \nu $ is defined as the
\textit{hierarchy index}.  This paper defines the host
\textit{pdf}'s $ p^{\nu} $ to deviate from the equilibrium state,
with increasing values of $ \nu $.

The case of incomplete constraints corresponds to scenarios where
the number of constraints (physical observable's) is less than the
dimension of the distribution. In a discrete setting, the MaxEnt
Lagrangian for a single host \textit{pdf} $
p^\nu(x_n)=p_n^\nu;n=1,...,N $ is [4]:
\begin{equation}\label{1}
L^{ME}  =  - \sum\limits_{n = 1}^N {p_n^\nu  \ln } p_n^\nu  + \sum\limits_{i = 0}^M {\sum\limits_{n = 1}^N {\lambda _i^\nu  p_n^\nu  \Theta _i \left( {x_n } \right)} },  \\
\end{equation}
$ \ M < N \ $, where the Lagrange multiplier (LM) $ \ \lambda _o^\nu
\ $ corresponds to the \textit{pdf} normalization condition $ \
\sum\limits_{n=1}^N {p_n^\nu } = 1 \ $. The LM's $ \ \lambda _i^\nu
;i = 1,...,M \ $ correspond to physical constraints of the form $ \
\sum\limits_{n=1}^N {p_n^\nu \Theta _i \left( {x_n } \right)}  =
d_i^\nu ;i = 1,...,M \ $. Here, $ \Theta _i \left( x_n \right);i =
1,...,M \ $ is an operator, and, $ \ d_i^\nu ;i=1,...,M $ are the
incomplete constraints. This paper employs geometric moment
constraints, i.e. $ \ \Theta _i \left( x_n \right) = x_n^i ;i =
1,...,M \ $.  Solution of (1) yields:
\begin{equation}\label{2}
\begin{array}{l}
 p_n^\nu   = \exp \left[ { - \sum\limits_{i = 0}^M {\lambda _i^\nu  {x_n^i }} } \right];n = 1,...,N, \\
 and, \\
 e^{\lambda _o^\nu  }  = \sum\limits_{n = 1}^N {e^{ - \lambda _i^\nu  x_n^i }} ;i = 1,...,M. \\
 \end{array}
\end{equation}
Solution of (2) for a given input set of LM's, is referred to as the
\textit{forward MaxEnt problem}. Inference of \textit{pdf}'s and the
concomitant LM's, given an input set of $ d_i^\nu ;i = 1,...,M $, is
referred to as the \textit{inverse MaxEnt problem}.

Statistical distributions inferred from incomplete constraints
possess eigenstructures that are ill-conditioned.  These demonstrate
extreme sensitivity to perturbations.   This sensitivity is
exploited to formulate a principled framework to encrypt/decrypt
code\footnote{In this paper, the terms \textit{code} and
\textit{covert information} are used interchangeably.}. The
\textit{null spaces }of these eigenstructures constitute an
``invisible reservoir'' into which covert information may be
projected. The projection is achieved without altering the host
statistical distributions.  The advantage of statistical encryption
is that the dimension of the code that can be encrypted increases
with the size of the statistical distribution. The projection of
code into a host statistical distribution is also a promising
candidate to implement steganography [5], and, related disciplines
in information hiding [6].

A recent study has treated the statistical encryption/decryption of
covert information, using the Fisher information as the measure of
uncertainty [7]. Qualitative distinctions vis-\'{a}-vis an
equivalent MaxEnt formulation [8] have been established. The
encoding strategies in [7] and [8] are \textit{independent} of the
\textit{pdf} of the host statistical distribution. Encoding covert
information into multiple \textit{pdf}'s allows code of dimension
greater than that of any single host \textit{pdf} to be selectively
encrypted into the multiple \textit{pdf}'s.  An example of this is
the encryption of multi-dimensional code defined in matrix form,
such as image data. In this case, each dimension of the code
(column/row vector of the matrix) may be selectively encoded into
individual components (single host \textit{pdf}'s) of a multiple
host \textit{pdf} model. Each single host \textit{pdf} is taken to
be one dimensional in the \textit{continuum}.

Conversely, the dimension of each host in a multiple \textit{pdf}
model may be chosen by the designer to be less than the dimension of
the code. This is tantamount to effecting a tradeoff between the
dimension of a single host \textit{pdf}, and, the number of hosts
comprising a multiple \textit{pdf} model. \textit{The success of
employing multiple pdf's to enhance the security of the covert
information, is critically contingent upon the encryption/decryption
strategy being dependent upon the pdf's of the statistical hosts}.
This feature permits the \textit{pdf} dependent statistical
encryption/decryption strategy to possess immense qualitative
flexibility, as compared with \textit{pdf} independent models [7,
8]. Numerical simulations demonstrate impressive quantitative
performance in securing covert information.

\section{Projection of the Covert Information}

Consider $ \ M \ $ constraints  $ \ d_1^\nu ,...,d_M^\nu \ $. In a
discrete setting, these are expectation values of a random variable
$ \ x_{i,n} ;n = 1,...,N \ $:
\begin{equation}\label{3}
\ d_i^\nu  = \sum\limits_{n = 1}^N {p_n^\nu x_{i,n} } ;i = 1,...,M.\
\end{equation}
Evoking the Dirac \textit{bra-ket} notation [9], the \textit{pdf} $
\ \left| p^\nu \right\rangle \in \Re ^N \ $ which is a column vector
(\textit{ket}), where $ \left| n \right\rangle ;n = 1,...,N \ $ is
the standard basis in $ \Re ^N $, is expressed as $ \left| p^\nu
\right\rangle  = \sum\limits_{n = 1}^N {\left| n \right\rangle
\left\langle {n}
 \mathrel{\left | {\vphantom {n p}}
 \right. \kern-\nulldelimiterspace}
 {p^\nu} \right\rangle }  = \sum\limits_{n = 1}^N {p_n^\nu \left| n \right\rangle }
 \ $.
Defining the column vector of observable's as $ \ \left| d^\nu
\right\rangle \in \Re ^{M + 1} \ $ with components $ \ d_1^\nu
,...,d_M^\nu,1 \ $, and, an operator $ A:\Re ^N  \to \Re ^{M + 1} $
given by $ A = \sum\limits_{n = 1}^N {\left| {x_n } \right\rangle
\left\langle n \right|} \ $. Defining vectors $ \ \left| {x_n }
\right\rangle \in \Re ^{M + 1} ;n = 1,...,N \ $ as the expansion $
\left| {x_n } \right\rangle  = \sum\limits_{i = 1}^{M + 1} {\left| i
\right\rangle } \left\langle {i}
 \mathrel{\left | {\vphantom {i {x_n }}}
 \right. \kern-\nulldelimiterspace}
 {{x_n }} \right\rangle  = \sum\limits_{i = 1}^{M + 1} {x_{i,n} } \left| i \right\rangle
 \ $, where $ \ i \ $ is a basis vector in $  \Re ^{M + 1}  $, (3)
may be expressed simply as:
\begin{equation}\label{7} \left| {d^\nu  } \right\rangle =
A\left| {p^\nu } \right\rangle ;A:\Re ^N  \to \Re ^{M + 1} .\
\end{equation}

The physical significance of the constraint operator $ \ A \ $ in
(4) is as follows. Inference of the \textit{pdf} from physical
observable's employing (2) is achieved by specifying $ \ \Theta _i
\left( {x_n } \right) = x_n^i \ $. Setting, $ \ x_n^i \to x_{i,n} ;i
= 1,...,M;n = 1,...,N \ $, the $ \ x_{i,n} \ $ constitute the
elements of the $ \ M \ $ rows and $ \ N \ $ columns of the operator
$ A^\nu $.  The unity element in $ \ \left| d^\nu \right\rangle \in
\Re ^{M + 1} \ $ enforces the normalization of $ \left| {p^\nu }
\right\rangle $.  Consequently, $ x_{M+1,n}=x_n^{M+1}=1;n=1,...,N $.

\textit{The operator $ \ A \ $ is independent of the host pdf}. This
qualitative deficiency may be rectified by defining:
\begin{equation}\label{3}
\ \tilde A^\nu = A + k^\nu\left| {d^\nu } \right\rangle \left\langle
I \right|. \
\end{equation}
Here,  $ \ \ \left\langle I \right| \ $ is a $\ 1 \times N \ $ row
vector (\textit{bra}), and, $ \ k^\nu \ \neq -1 $ is a constant
parameter introduced to adjust the condition number of $  \ \tilde
A^\nu \ \ $, and hence its sensitivity to perturbations.  In (5),
dependence upon the host \textit{pdf} is "injected" into the
operator $  \ \tilde A^\nu \ \ $ by the incorporation of $ \ \left|
{d^\nu } \right\rangle \ $.  Specifically, each element of the
\textit{ket} $ \ \left| {d^\nu } \right\rangle \ $ is defined by $
d_i^\nu= \sum\limits_{n = 1}^N {p_n^\nu  } x_n^i ;i = 1,...,M  $.

Re-defining (4) in terms of (5) yields: $ \ \tilde A^\nu  \left|
{p^\nu } \right\rangle = \left| {d^\nu } \right\rangle  +
\left\langle {k^\nu\left| {d^\nu } \right\rangle \left\langle I
\right|} \right\rangle _\nu \ $, where $ \ \left\langle \bullet
\right\rangle _\nu \ $ signifies the expectation evaluated at the
\textit{hierarchy index} $ \nu \ $. Expanding $ \ \left\langle
{k^\nu\left| {d^\nu } \right\rangle \left\langle I \right|}
\right\rangle _\nu = k^\nu\left| {d^\nu } \right\rangle \left\langle
I \right|\left. {p^\nu  } \right\rangle \ $, and evoking the
\textit{pdf} normalization: $ \ \left\langle I \right|\left. {p^\nu
} \right\rangle  = 1 \ $, yields:
\begin{equation}\label{4}
\left| {\tilde d^\nu  } \right\rangle  = \left( {k^\nu + 1}
\right)\left| {d^\nu  } \right\rangle  = \tilde A^\nu \left| {p^\nu
} \right\rangle ;\tilde A^\nu  :\Re ^N  \to \Re ^{M + 1} .\
\end{equation}

The operator $ \ \tilde A^\nu \ \ $is ill-conditioned and
rectangular.  Thus, (6) becomes:
\begin{equation}\label{3}
\left| {p^\nu  } \right\rangle  = \left( {\tilde A^\nu  } \right)^{
- 1} \left| {\tilde d^\nu  } \right\rangle  + \left| {p^{\nu '} }
\right\rangle, \
\end{equation}
where, $ \ {\left( {\tilde A^\nu  } \right)^{ - 1} } \ $ is the
pseudo-inverse [10] of $ \ \tilde A^\nu \ \ $, and lies in $ \
range\left( {\tilde A^\nu  } \right) \ $. All necessary data
dependent information resides in $ \left( {\tilde A^\nu } \right)^{
- 1} \left| {\tilde d^\nu  } \right\rangle $.

The \textit{null space} term in (7) is of particular importance
since the code is encoded into it via unitary projections.  Here, $
\ \left| {p^{\nu '} } \right\rangle  \in null\left( {\tilde A^\nu  }
\right) \ $ is explicitly data independent\footnote{In this paper, $
null\left( {} \right) $ signifies the \textit{null space} of an
ill-conditioned operator, unless explicitly specified as being the $
MATLAB^{\circledR} $ routine to calculate the normalized basis of an
ill-conditioned operator (eg. Section 3.1).}. However, it is
critically dependent on the solution methodology employed to solve
(7).  To define the unitary projections of the embedded code, the
operator $ \ G^\nu = \tilde A^{\nu \dag} \tilde A^\nu \ $ is
introduced. Here, $ \tilde A^{\nu \dag} \ $ is the conjugate
transpose of $ \tilde A^\nu \ $. For real matrices, $ \ \tilde
A^{\nu \dag } = \tilde A^{\nu T} \ $, where $ \ \tilde A^{\nu T} \ $
is the transpose of $ \ \tilde A^\nu \ $.

It is legitimate to project the covert information into $ \
null\left( {\tilde A^\nu  } \right) \ $ instead of $ \ null\left(
{G^\nu  } \right) \ $ [7]. The ``loss of information" caused by
floating point errors could make the evaluation of  $ \ null\left(
{G^\nu } \right) \ $ prohibitively unstable for many applications.
Specifically, this operation squares the condition number, resulting
in the large singular values being increased and the small singular
values decreased.  This instability is used to the advantage of the
designer to increase the security of the covert information.

Assuming the availability of $ \tilde A^\nu \ $ and $ \left| p^\nu
\right\rangle $, the normalized eigenvectors corresponding to the
eigenvalues in the \textit{null space} of $ \ G^\nu $ having value
zero (\textit{zero eigenvalues}) is: $ \left| {\eta _n^\nu }
\right\rangle ;n = 1,...,N - (M + 1) $. The $ \left| {\eta _n^\nu }
\right\rangle $ are hereafter referred to as the basis of $ \
null\left( {G^\nu  } \right) $.  The unitary decryption $ \ \hat
U_{dec}^\nu \ :\Re ^N \to \Re ^{N - (M + 1)} \ $ and encryption
operators $ \ \hat U_{enc}^\nu \ :\Re ^{N - (M + 1)} \to \Re ^N \ $
operators for each $ \nu $ are:
\begin{equation}\label{2}
\begin{array}{l}
 \hat U_{dec}^\nu = \sum\limits_{n = 1}^{N - M - 1} {\left| n
\right\rangle \left\langle {\eta _n^\nu  } \right|} , \\
and, \\
\hat U_{enc}^\nu   = \hat U_{dec}^{\nu \dag }  = \sum\limits_{n =
1}^{N - M - 1} {\left| {\eta _n^\nu  } \right\rangle \left\langle n
\right|}, \\
 \end{array}
\end{equation}
respectively.  Note that $ \ \hat U_{dec}^\nu   \bullet \hat
U_{enc}^\nu   = I \ $,  where $ \ I \ $ is the identity operator.

Given a code $ \left| q^\nu  \right\rangle \in \Re ^{N - (M + 1)} $
to be encrypted in a host \textit{pdf} having \textit{hierarchy
index} $ \nu $ , the $ \ N - \left( {M + 1} \right) \ $ components
are given by $ \left\langle {n}
 \mathrel{\left | {\vphantom {n {q^\nu  }}}
 \right. \kern-\nulldelimiterspace}
 {{q^\nu  }} \right\rangle  = q_n^\nu  ;n = 1,...,N - \left( {M + 1} \right)
\ $.  The \textit{pdf} of the embedded code is:
\begin{equation}\label{3}
\left| {p_c^\nu  } \right\rangle  = \hat U_{enc}^\nu \left| {q^\nu }
\right\rangle  = \sum\limits_{n = 1}^{N - M - 1} {\left| {\eta
_n^\nu  } \right\rangle \left\langle {n}
 \mathrel{\left | {\vphantom {n {q^\nu  }}}
 \right. \kern-\nulldelimiterspace}
 {{q^\nu  }} \right\rangle }. \
\end{equation}

The total \textit{pdf} comprising the host \textit{pdf} and the
\textit{pdf} of the code is:
\begin{equation}\label{7}
\left| {\tilde p^\nu  } \right\rangle  = \left| {p^\nu  }
\right\rangle  + \left| {p_c^\nu  } \right\rangle. \
\end{equation}
Since $ \ \left| {p_c^\nu  } \right\rangle  \in null\left( {\tilde
A^\nu  } \right) \ $, $ \ \tilde A^\nu \left| {p_c^\nu }
\right\rangle = 0 \ $. In the decryption stage, the reconstructed
host \textit{pdf}'s: $ \left| {p_r^\nu } \right\rangle  $, are first
obtained. The \textit{pdf}'s of the embedded code is recovered from:
\begin{equation}\label{3}
\left| {p_{rc}^\nu  } \right\rangle  = \left| {\tilde {p^\nu }}
\right\rangle  - \left| {p_r^\nu } \right\rangle. \
\end{equation}

The encrypted code is recovered by the operation:
\begin{equation}\label{3}
\left| q_r^\nu \right\rangle  = \hat U_{dec}^\nu \left| {p_{rc}^\nu
} \right\rangle = \sum\limits_{n = 1}^{N - M - 1} {\left| n
\right\rangle \left\langle {{\eta _n^\nu }}
 \mathrel{\left | {\vphantom {{\eta _n } {p_c^\nu  }}}
 \right. \kern-\nulldelimiterspace}
 {{p_{rc}^\nu }} \right\rangle }. \
\end{equation}

The above theory does not, by itself, constitute the strategy to
encrypt/decrypt code. This is achieved in two manners,i.e,
\textit{symmetric} and \textit{asymmetric} cryptographic strategies
[11,12]. Before proceeding further, the concept of a key in the
encryption of covert information is briefly explained.  The
distribution of keys is an issue of primary concern in cryptography
and allied disciplines. A key may be a program, a number, or a
string of numbers that enables the legitimate recipient of the
message (decrypter) to access the covert information. In
cryptography, a secret, shared, or private key is an
encryption/decryption key known only to the entities that exchange
secret messages.

In traditional secret key cryptography, a key would be shared by the
communicators so that each could encrypt and decrypt messages. The
risk in this system is that if either party loses the key or it is
stolen, the system is broken. Secret key cryptography is also
susceptible to a number of malicious attacks, the most common being
the \textit{plaintext attack} [11,12]. By definition, a
\textit{plaintext attack} is one where the prior messages have
intercepted and decrypted in order to decrypt other messages. A more
recent alternative is to use a combination of public and private
keys. In this system, a public key is used together with a secret
key.  The RSA protocol [1] is a prominent example of a public key
infrastructure (PKI). A PKI often employs a \textit{key ring
strategy}.  Specifically, one key is kept secret while the others
are made public.  PKI is the preferred approach on the Internet. The
secret key system is sometimes known as \textit{symmetric}
cryptography and the public key system as \textit{asymmetric}
cryptography.

In this model, an operator $ \tilde G^\nu $ is formed by perturbing
select elements of $ \ G^\nu \ $ by $ \delta G_{i,j}^\nu $. In
\textit{symmetric} cryptography, only a single element of  $ \ G^\nu
\ $ is perturbed. The security of the code may be ensured by
adopting an \textit{asymmetric} cryptographic strategy. Here, more
than one element of $ \ G^\nu \ $ is perturbed. Each $ \delta
G_{i,j}^\nu > \delta ^\nu $, a threshold.  The extreme sensitivity
to perturbations of $ \ G^\nu \ $ causes the eigenstructure of $
{\tilde G^\nu = \ G^\nu + \delta G_{i,j}^\nu } \ $ to substantially
differ from that of $ {G^\nu } $, This assertion is valid even for
infinitesimal perturbations $ \delta G_{i,j}^\nu $. To distinguish
operations involving the ill-conditioned operators $ \tilde G^\nu $,
the following change of notation is effected in (9)-(12): $ \left|
{q^\nu } \right\rangle \to \left| {\tilde q^\nu } \right\rangle $,$
\left| {\eta_n^\nu  } \right\rangle \to \left| {\tilde \eta_n^\nu  }
\right\rangle  $, $ \left| {p_c^\nu  } \right\rangle \to \left|
{\tilde p_c^\nu  } \right\rangle  $, $ \left| {\tilde p^\nu }
\right\rangle \to \left| {\tilde p_{pert}^\nu } \right\rangle  $,$
\left| {p_{rc}^\nu  } \right\rangle  \to \left| {\tilde p_{rc}^\nu }
\right\rangle  $, and, $ \left| {q_{r}^\nu  } \right\rangle  \to
\left| {\tilde q_{r}^\nu } \right\rangle  $. The $ \left| {\tilde
\eta_n^\nu  } \right\rangle  $ are hereafter referred to as the
basis of $ null\left( {\tilde G^\nu  } \right) \ $.

Determination of the threshold is a vital task in defining the
cryptographic keys in this model.  This is accomplished by the
designer (encrypter) who performs a simultaneous
encryption/decryption without effecting perturbations to the
operator $ G^\nu $.  Specifically, using (2), the host
\textit{pdf}'s are inferred by solving an \textit{inverse MaxEnt
problem}. The code $ \ \left| {q^\nu } \right\rangle \ $ having
dimension $ N-(M+1) $ is formed.  The designer implements (9)-(12)
for each \textit{hierarchy index} $ \nu $.  The threshold for the
cryptographic key/keys is: $  \delta ^\nu = \left\| {\left| {q^\nu }
\right\rangle - \left| {q_r^\nu } \right\rangle } \right\|  $.

\section{Implementation of the Encryption/Decryption Strategy}
The process of encryption occurs after the host \textit{pdf}'s have
been inferred from incomplete constraints for \textit{multiple
pdf}'s. This procedure is detailed in Section 4. The terminology in
cryptography and allied disciplines refers to two communicating
parties as Alice and Bob, and, an eavesdropper as Eve. In this
study, the author performs the role of both Alice and Bob by
implementing the encryption on an IBM RS-6000 workstation cluster
and, the decryption on an IBM Thinkpad running $ MATLAB^{\circledR}
$ v 7.01.

Herein, the implementation of the encryption/decryption strategy by
effecting perturbations $ \delta G_{i,j}^\nu $ to the operator $ \
G^\nu \ $ is described. The procedures are to be implemented for
each \textit{hierarchy index} $ \nu $. This Section presents the
implementation of the encryption/decryption strategies in point
form, for the sake of clarity and brevity.

\subsection{Encryption}  $ \left( {i.} \right) $   The host
\textit{pdf}  $ \left| {p^\nu } \right\rangle $ is inferred from
incomplete constraints by solving (2) as an \textit{inverse MaxEnt
problem}.  $ \left( {ii.} \right) $ The constraint operators $
\tilde A^\nu $ and $ G^\nu $ are evaluated formed, for an
\textit{a-priori} specified value of the parameter $ k^\nu $, from
(5). The operator $ \tilde G^\nu $ is formed by perturbing one or
more elements of $ \ G^\nu \ $ by $ \delta G_{i,j}^\nu
> \delta ^\nu $, the cryptographic key/keys.  $ \left( {iii.} \right) $
The basis $ \left| {\tilde \eta _n^\nu  } \right\rangle ;n = 1,...,N
- (M + 1) \ $, are evaluated by operating on $ {\tilde G^\nu } $
with the $ MATLAB^{\circledR} $ routine $ \ null\left( \bullet
\right) \ $ that employs SVD, or with an equivalent routine. $
\left( {iv.} \right) $ The code $ \left| {\tilde q^\nu }
\right\rangle $ is generated, and, is encoded into the \textit{null
space} of $ \tilde G^\nu $ using (9). $ \left( {v.} \right) $ The
total \textit{pdf} $  \left| {\tilde p_{pert}^\nu } \right\rangle $
is obtained using (10).

\subsection{Transmission}

The statistical encryption model provides two separate manners in
which information may be transferred from the encrypter to the
decrypter, via a \textit{public channel}.  The first mode is to
transmit the constraint operators $ \tilde A^\nu $ and the total
\textit{pdf}'s $  \left| {\tilde p_{pert}^\nu  } \right\rangle $. An
alternate mode is to transmit the LM's obtained on solving the
\textit{inverse MaxEnt problem}(Section 3.1), and, the total
\textit{pdf}'s $ \left| {\tilde p_{pert}^\nu  } \right\rangle $.
Owing to the large dimensions of the constraint operators $ \tilde
A^\nu $, the latter transmission strategy is more attractive.  The
values of parameters $ k^\nu $ for each \textit{hierarchy index},
and, the cryptography key/keys are transmitted through a
\textit{secure/covert channel}. The key/keys are labeled in order to
identify the elements of the operator $ \ G^\nu $ that are
perturbed. In the case of \textit{asymmetric} cryptography, some of
the keys may be declared public, while keeping the remainder
private.
\subsection{Decryption}

$ \left( {vi.} \right) $   The legitimate receiver recovers the
key/keys $ \delta G_{i,j}^\nu $ and the parameter $ k^\nu $ from the
\textit{covert channel}.  $ \left( {vii.} \right) $   The operators
$ \tilde A^\nu  \ $, $ \ G^\nu \ $, and, $ \tilde G^\nu  \ $ are
constructed. $ \left( {viii.} \right) $  The host \textit{pdf} may
be recovered in two distinct manners, depending upon the
transmission strategy employed. Note that both methods to
reconstruct the host \textit{pdf} require the total \textit{pdf} $
{\tilde p_{pert}^\nu  } $ to be provided by the encrypter.  First,
the scaled incomplete constraints, defined in (6), are obtained by
solving $ \ \left\langle i \right|\tilde A^\nu \left| {\tilde
p_{pert}^\nu  } \right\rangle \ = \ \left| {\tilde d^\nu }
\right\rangle \ $.  Here, $ \ i \ $ is a basis vector in $  \Re ^{M
+ 1} \ $.  This procedure is possible because $ \left| {\tilde
p_c^\nu  } \right\rangle  \in null\left( {\tilde A^\nu  } \right) $.
Thus, $ \ \tilde A^\nu \left| {\tilde p_c^\nu } \right\rangle = 0 \
$. The host \textit{pdf} are then computed for each
\textit{hierarchy index} by solving the \textit{inverse MaxEnt
problem}, using the re-scaled set of incomplete constraints.
Alternatively, the host \textit{pdf} may be obtained by solving (2)
as a \textit{forward MaxEnt problem}, given the values of the LM's $
\lambda _i^\nu \ ; i=1,...,M $ obtained from the \textit{inverse
MaxEnt problem} (Section 3.1). Both methods allow the reconstructed
host \textit{pdf}'s $ \left| {p_r^\nu } \right\rangle $ to be
obtained with a high degree of precision.  $ \left( {ix.} \right) $
The reconstructed code \textit{pdf} $ \left| {\tilde p_{rc}^\nu  }
\right\rangle $ is recovered using (11).  $ \left( {x.} \right) $
The reconstructed code $ \left| {\tilde q_r^\nu } \right\rangle $ is
obtained using (12).

It is important to note that the success of the
encryption/decryption strategy is critically dependent upon the
exact compatibility of software available to the encrypter and
decrypter. Of special importance is the compatibility of the
routines to calculate the basis $ \left| {\tilde \eta _n^\nu }
\right\rangle $.

\section{Numerical Simulations}
This Section provides numerical simulations to analyze the theory
and implementation of the statistical encryption/decryption
strategy, presented in Section's 2. and 3., respectively.  To
demonstrate the efficacy of the theory presented in this paper, it
is judicious to compare the \textit{pdf} dependent model with a
\textit{pdf} independent model [8].  These are characterized by the
constraint operators $ \tilde A^\nu $ (described in (5) and (6)),
and, $ \ A \ $ (described in (4)), respectively. The following
numerical studies perform the encryption/decryption strategy for the
case of \textit{asymmetric} cryptography.  Section 4.1 demonstrates
the inference of the host \textit{pdf}'s from incomplete
constraints, using an \textit{inverse MaxEnt} procedure.

The comparative analysis between the \textit{pdf} dependent model
and the \textit{pdf} independent model is accomplished using two
separate metrics, that define the security of the covert
information. Host \textit{pdf}'s that deviate further from the
equilibrium state often possess a constraint operator $ \tilde A^\nu
$ having a higher condition number, as compared with \textit{pdf}'s
that are closer to the equilibrium state.  Increasing the condition
number of the constraint operator $ \tilde A^\nu  $ (or, $ A $)
represents one way of securing the integrity of the covert
information.  The condition numbers of the constraint operators are
obtained using the $ MATLAB^{\circledR} $ routine $ \ cond\left(
\bullet \right) $. This provides a measure of the sensitivity of $
null\left( {G} \right) $ and $ null\left( {G^\nu } \right) $ to
perturbations.

In the \textit{pdf} dependent model, the security of the code is
enhanced by the increased sensitivity of $ null\left( {G^\nu }
\right) $ to perturbations of select elements of the operator $
{G^\nu } $, resulting in the operator $ {\tilde G^\nu } $ with
\textit{null space} denoted by $ null\left( {\tilde G^\nu } \right)
$. Within the framework of this model, a more relevant metric of the
extreme sensitivity of $ null\left( {G^\nu } \right) $ to
perturbations, brought about by the introduction of the
cryptographic keys $ \delta G_{i,j}^\nu $, is the distortion of the
code \textit{pdf} $ \left| {p_c^\nu } \right\rangle  $. Here, $
\left| {p_c^\nu } \right\rangle $ is evaluated from (9), using $
\eta_n^\nu $ (the basis of $ null\left( {G^\nu } \right) $). The
distorted code \textit{pdf} is $ \left| {\tilde p_c^\nu }
\right\rangle $, which is calculated from (9) using $ \tilde
\eta_n^\nu $ (the basis of $ null\left( {\tilde G^\nu } \right) $).
Note that the codes $ \left| {\tilde q^\nu } \right\rangle $ and $
\left| {q^\nu } \right\rangle $ are identical \textit{kets}.

For the \textit{pdf} dependent model, the \textit{RMS error of
encryption} between $ \left| {\tilde p_c^\nu } \right\rangle $ and $
\left| {p_c^\nu } \right\rangle  $ is defined as: $ RMS_{enc}^{\nu}
= \frac{{\left\| {err_{enc}^{\nu} } \right\|}}{{\sqrt {length\left(
{err_{enc}^{\nu} } \right)} }} $.  Here, $ \left\| {err_{enc}^{\nu}
} \right\| = \left\| {\left( {\left| {\tilde p_c^{\nu} }
\right\rangle - \left| {p_c^{\nu} } \right\rangle } \right)}
\right\| \ $, and, $ length\left( {err_{enc}^{\nu} } \right) $ is
the dimension of $ {\left( {\left| {\tilde p_c^{\nu} } \right\rangle
- \left| {p_c^{\nu} } \right\rangle } \right)} $. Section 4.3 will
exemplify the utility of $ RMS_{enc}^\nu $ in resolving a dichotomy.
Specifically, it will be demonstrated that a high value of $
RMS_{enc}^\nu $ is the reason for host \textit{pdf}'s possessing
constraint operators with lower condition numbers, sometimes
providing a greater degree of security to the encoded covert
information, than host \textit{pdf}'s possessing constraint
operators with higher condition numbers.

Another measure of the security of the covert information, that is
operationally advantageous, is the RMS error of the difference
between the encrypted code, and, the code reconstructed without the
cryptographic keys.  For the \textit{pdf} dependent model, the
\textit{RMS error of reconstruction} is: $ RMS_{recon}^{\nu}  =
\frac{{\left\| {err_{recon}^{\nu} } \right\|}}{{\sqrt {length\left(
{err_{recon}^{\nu} } \right)} }} $. Here, $ \ \left\|
{err_{recon}^{\nu} } \right\| = \left\| {\left( {\left| {\tilde
q^{\nu} } \right\rangle - \left| {\tilde q_{r1}^{\nu} }
\right\rangle } \right)} \right\| \ $, and, $ length\left(
{err_{recon}^{\nu} } \right) $ is the dimension of $ {\left( {\left|
{\tilde q^{\nu} } \right\rangle - \left| {\tilde q_{r1}^{\nu} }
\right\rangle } \right)} $.

The rationale for evaluating the $ RMS_{recon}^{\nu} $ is to obtain
a measure of the error of reconstruction by an \textit{unauthorized
eavesdropper (Eve)}, who does not possess the correct cryptographic
keys $  \delta G_{i,j}^\nu $.  Eve is, however, assumed to be in
possession of the reconstructed code \textit{pdf} $ \left| {\tilde
p_{rc}^\nu } \right\rangle  $.
 Such an attack may be simulated
 by choosing the wrong set of  keys, and, incorrectly constructing the perturbed operators
 $ \tilde G^\nu $ and $ \tilde G \ $.  Alternatively, an analogous scenario may be simulated by assuming
that the reconstruction is performed without keys (\textit{sans}
keys).  In this case, the reconstructed code without keys is: $
\left| {\tilde q_{r1}^\nu  } \right\rangle = \sum\limits_{n = 1}^{N
- M - 1} {\left| n \right\rangle \left\langle {{\eta _n^\nu }}
 \mathrel{\left | {\vphantom {{\eta _n^\nu} {\tilde p_{rc}^\nu  }}}
 \right. \kern-\nulldelimiterspace}
 {{\tilde p_{rc}^\nu  }} \right\rangle } $.  Here, $ \eta_n^\nu $ is the basis of
$ null\left( {G^\nu } \right) $.  Section 4.2 presents the comparative analysis using the performance
metrics described above.  Analysis of the results is provided in
Section 4.3.

\subsection{Inference of the host pdf's}
Host \textit{pdf }column vectors (\textit{kets}) corresponding to
\textit{hierarchy indices} $ \nu=1 $ and $ \nu =2 $, each of
dimension 401, are independently inferred in the event space $ \
\left[ { - 1,1} \right] \ $. The incomplete constraints are the
first four moments of the random variable $ x_n ;n=1,...,N $. From
(2), one obtains:
\begin{equation}\label{3}
 d_i^\nu   = \frac{{\sum\limits_{n = 1}^{N = 401} {x_n^i e^{ - \lambda _1^\nu  x_n  - \lambda _2^\nu  x_n^2  - \lambda _3^\nu  x_n^3  - \lambda _4^\nu  x_n^4 } } }}{{\sum\limits_{n = 1}^{N = 401} {e^{ - \lambda _1^\nu  x_n  - \lambda _2^\nu  x_n^2  - \lambda _3^\nu  x_n^3  - \lambda _4^\nu  x_n^4 } } }}; i = 1,...,M=4 \
\end{equation}
Here, (13) is solved as an \textit{inverse MaxEnt problem} for two
different sets of incomplete constraints (corresponding to each
\textit{hierarchy index}), provided as input values. These are $
d_1^{\nu = 1} = {\rm - 0}{\rm .0224,}d_2^{\nu  = 1} = {\rm 0}{\rm
.1048,}d_3^{\nu = 1} = {\rm  - 0}{\rm .0124,}d_4^{\nu  = 1} = {\rm
0}{\rm .0284} $, and, $ d_1^{\nu  = 2}  = 0.1{\rm ,}d_2^{\nu = 2} =
{\rm 0}{\rm .3,}d_3^{\nu  = 2}  = 0.1{\rm ,}d_4^{\nu  = 2}  = {\rm
0}{\rm .15} $, respectively. Note that the values of $ d_i^{\nu=1}
$, are taken to be the same as those in [8].

Each \textit{hierarchy index} yields $ M+1 $ constraint equations,
that are solved using a Newton-Raphson procedure. Note that the $
(M+1)^{st} $ constraint equation follows from the \textit{pdf}
normalization condition. The values of the LM's are found to be $ \
\lambda _1^{\nu = 1} = - {\rm 0}{\rm .30435,}\lambda _2^{\nu = 1} =
{\rm 2}{\rm .99664,}\lambda _3^{\nu = 1}  = {\rm 4}{\rm
.85637,}d_4^{\nu = 1} = {\rm 3}{\rm .81359} \ $, and, $ \ \lambda
_1^{\nu  = 2}  = {\rm 1}{\rm .74906,}\lambda _2^{\nu  = 2} = - {\rm
5}{\rm .09475,}\lambda _3^{\nu  = 2}  =  - {\rm 4}{\rm
.8568,}\lambda _4^{\nu  = 2}  = {\rm 8}{\rm .48337} \ $,
respectively.

In this example, each constraint operator $ \tilde A^\nu  \ $ is of
dimension $ \ 5 \times 401 \ $, and, the number of basis $ \tilde
\eta _n^{\nu=1,2} $ is 396, for each \textit{hierarchy index}.
Figure 1 depicts the two host \textit{pdf}'s. The case with
\textit{hierarchy index} $ \nu=1 $ is the single peaked
\textit{pdf}, while the case with \textit{hierarchy index} $ \nu=2 $
is the double peaked \textit{pdf}.

\begin{figure}[thpb]
      \centering
       \includegraphics[scale=0.7]{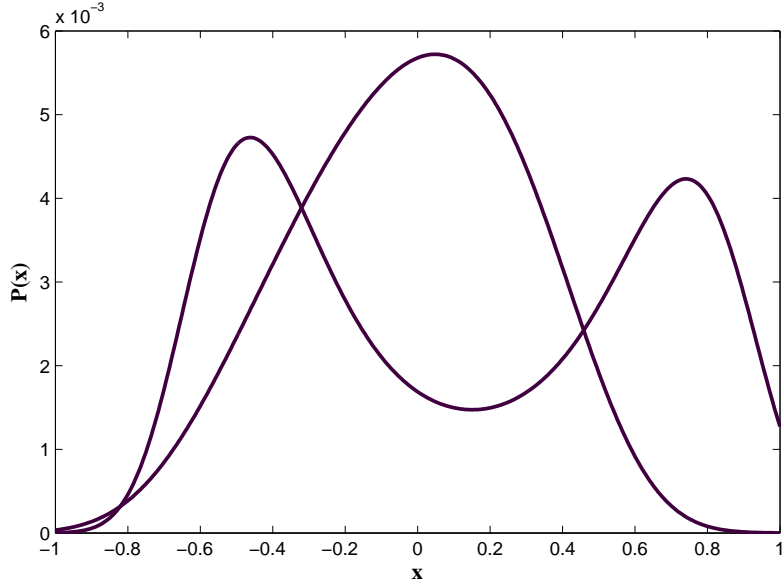}
    \caption{Host \textit{pdf}'s inferred from incomplete constraints for $ \nu=1 $ and $ \nu=2 $}
    \label{fig:Quadratic}
\end{figure}
\subsection{Comparative analysis}
The $ MATLAB^{\circledR} $ random number generator $ \ rand\left(
\bullet \right) \ $ is evoked to generate code in $ \ \left[ {0,1}
\right] \ $.  In order to establish a degree of uniformity in the
comparisons, two identical \textit{kets} of the code, each having
dimension $ N-(M+1) $ i.e. $ 396 $, are created for projection into
the \textit{null spaces} of the perturbed operators $ \tilde G^\nu $
and $ \tilde G $, respectively. This "emulates" the
\textit{selective} projection of a code comprising of a single
\textit{ket }of dimension $ 792  $, into $ null\left( {\tilde
G^{\nu=1,2} } \right) $ and $ null\left( {\tilde G}  \right) $, for
each \textit{hierarchy index}.

A further measure of uniformity in the comparative analysis is
sought by specifying the perturbations to both $ \ G \ $ and $ \
G^\nu \ $ as $ \ \delta G_{1,3}  = \delta \tilde G_{1,3}^{\nu  =
1,2} = 3.0e - 013 \ $ (first row, third column) and $ \ \delta
G_{2,7}  = \delta \tilde G_{2,7}^{\nu = 1,2}  = 7.0e - 013 \ $
(second row, seventh column), respectively. All numerical examples
in this paper have a threshold for perturbations $ \ \delta ^\nu
\sim 4.4e - 014 \ $.

Simulations for the case of the \textit{pdf} independent model are
described in Section 4.2.1. Two distinct case studies for the
\textit{pdf} dependent model are described in Sections 4.2.2 and
4.2.3, respectively. The results are presented in Table 1 through
Table 3. Therein, sample values of the original code, the code
reconstructed \textit{with} the keys, and, the code reconstructed
\textit{without} the keys are presented. These comprise the $ \
1^{st} ,75^{th} ,177^{th}, 296^{th}, and,395^{th} \ $ elements of
the respective arrays (\textit{kets}). As is evident from the
numerical simulations, the reconstructed code \textit{with} the keys
is exactly similar to the original code. The corresponding $
RMS_{recon}^\nu $ is zero.  On the other hand, the code
reconstructed without the keys bears no resemblance to the original
code.

\subsubsection{Pdf independent model} Here, $ \ cond\left( A \right)
= {\rm 18}{\rm .80458} \ $, $ RMS_{enc}^{\nu=1,2}=0.76998 $, and, $
RMS_{recon}^{\nu=1,2} = {\rm 0}{\rm .77483} \ $.  The values of the
encrypted and reconstructed codes are independent of the
\textit{hierarchy index}. These results are consistent with
expectations since the nature of the host \textit{pdf} is not
reflected in the constraint operator $ A $. Select code values are
presented in Table 1.

\begin{table}
\caption{Select code values for the \textit{pdf} independent model.
 $ \left|
{q^\nu } \right\rangle $ - the original code,$ \left| {q_r^\nu }
\right\rangle $ - the code reconstructed \textit{with} the keys,
and, $ \left| {q_{r1}^\nu } \right\rangle $ - the code reconstructed
\textit{without} the keys. } \label{table_example}

\begin{tabular}{c c c}
\hline $ \ \left| {q^{\nu=1,2}  } \right\rangle \ $
 & $ \ \left| {q_r^{\nu=1,2} } \right\rangle \ $ & $ \ \left| {q_{r1}^{\nu=1,2} } \right\rangle \  $ \\
\hline
0.23813682639005 &  0.23813682639005 &    -0.09239755885631 \\
0.69913526160795 &  0.69913526160795 &   0.20008072567388 \\
0.27379424177629 & 0.27379424177629 &     -0.16469110441464 \\
0.17686701421432 & 0.17686701421432 &   0.92507582836680 \\
0.20288628732009 & 0.20288628732009 &   -0.51825561367528 \\
\hline
\end{tabular}
\end{table}

\subsubsection{Pdf dependent model : case study 1} The parameters are
$ k^{\nu=1}=0.065 $, and, $ k^{\nu=2}=0.09 $. Here, $ cond\left(
{\tilde A^{\nu = 1} } \right) = 19.{\rm 87136} $, $
RMS_{enc}^{\nu=1} = {\rm 0}{\rm .79533} $, and, $
RMS_{recon}^{\nu=1} = {\rm 0}{\rm .80034} $.  Further, $ cond\left(
{\tilde A^{\nu = 2} } \right)\\ = {\rm 20}{\rm .42975} $, $
RMS_{enc}^{\nu=2} = {\rm 0}{\rm .82570} $, and, $
RMS_{recon}^{\nu=2} = {\rm 0}{\rm .83090} $. This case represents a
noticeable increase in the level of security of the code, and
flexibility of the theory, as compared with the \textit{pdf}
independent model. Select code values are presented in Table 2.
\begin{table}
\caption{Select code values for the \textit{pdf} dependent
model-case study 1.
 $ \left| {\tilde q^\nu }
\right\rangle $ - the original code,$ \left| {\tilde q_r^\nu }
\right\rangle $ - the code reconstructed \textit{with} the keys,
and, $ \left| {\tilde q_{r1}^\nu } \right\rangle $ - the code
reconstructed \textit{without} the keys.} \label{table example}
\begin{tabular}{c c c c}
\hline $ \left|{\tilde q^{\nu=1,2}  } \right\rangle $
 & $ \left| {\tilde q_r^{\nu=1,2} } \right\rangle $ & $ \left| {\tilde q_{r1}^{\nu=1}  } \right\rangle $ & $ \left| {\tilde q_{r1}^{\nu=2}  } \right\rangle $ \\
\hline
0.23813682639005 &  0.23813682639005 &    -0.08237185729405 & 0.40428740297403 \\
0.69913526160795 &  0.69913526160795 &   -0.01133471359936 & 0.30190786147110 \\
0.27379424177629 & 0.27379424177629 &      -0.33080751552396 & -0.29261640748833 \\
0.17686701421432 & 0.17686701421432 &     -1.18947579992697 & 0.53834101226185 \\
0.20288628732009 & 0.20288628732009 &      0.54528163427934 & -0.10688384159491 \\
\hline
\end{tabular}
\end{table}

\subsubsection{Pdf dependent model :case study 2 - a study in
contrast}

The parameters are $ k^{\nu=1}=-0.03 $, and, $ k^{\nu=2}=-0.5  $.
Here, $ cond\left( {\tilde A^{\nu = 1} } \right) = 18.{\rm 31597} $,
$ RMS_{enc}^{\nu=1} = 0.81620 $ and, $ RMS_{recon}^{\nu=1} = {\rm
0}{\rm .82134} $.  Further, $ cond\left( {\tilde A^{\nu = 2} }
\right) \\= {\rm 11}{\rm .97782} $, $ RMS_{enc}^{\nu=2} = {\rm
0}{\rm .84434} \ $, and, $ RMS_{recon}^{\nu=2} = {\rm 0}{\rm .84965}
$. This case represents a noticeable increase in the level of
security of the code in terms of $ RMS_{recon}^{\nu=1,2} $, the RMS
error of reconstruction, as compared with the results in Sections
4.2.1 and 4.2.2, respectively. Select code values are presented in
Table 3.

\begin{table}
\caption{Select code values for the \textit{pdf} dependent
model-case study 2.
 $ \left| {\tilde q^\nu }
\right\rangle $ - the original code,$ \left| {\tilde q_r^\nu }
\right\rangle $ - the code reconstructed \textit{with} the keys,
and, $ \left| {\tilde q_{r1}^\nu } \right\rangle $ - the code
reconstructed \textit{without} the keys.} \label{table_example}
\begin{tabular}{c c c c} \hline $ \left|{\tilde q^{\nu=1,2}  }
\right\rangle $
 & $ \left| {\tilde q_r^{\nu=1,2} } \right\rangle $ & $ \left| {\tilde q_{r1}^{\nu=1}  } \right\rangle $ & $ \left| {\tilde q_{r1}^{\nu=2}  } \right\rangle $ \\
\hline
0.23813682639005 &  0.23813682639005 &    0.41205157405440  & -1.47600706971518 \\
0.69913526160795 &  0.69913526160795 &   0.96359310993344 & -0.12406066610739 \\
0.27379424177629 & 0.27379424177629 &     -0.03625400789579 & 0.71289692833557 \\
0.17686701421432 & 0.17686701421432 &   -0.94767053576282 & -0.37638356088402 \\
0.20288628732009 & 0.20288628732009 &    -0.87540038881240 & 0.57110616844674 \\
\hline
\end{tabular}
\end{table}

\subsection{Analysis of results}
The case presented in Section 4.2.2 follows expectations.
Specifically, the constraint operators $ \tilde A^{\nu=1,2}  \ $
possess larger condition numbers than the constraint operator $ \ A
\ $ of the \textit{pdf} independent model, presented in Section
4.2.1. Further, the $ RMS_{enc}^{\nu=1,2} $ and $
RMS_{recon}^{\nu=1,2} $ in the \textit{pdf} dependent model are
greater those for the \textit{pdf} independent model. This implies
that the \textit{pdf} dependent model provides greater security to
the covert information, than the \textit{pdf} independent model. The
case presented in Section 4.2.3 represents an interesting scenario,
which poses a dichotomy of sorts.

Conventional logic would expect an increase in the condition numbers
of $ \tilde A^{\nu=1,2}  \ $ to coincide with an increase in the
sensitivity of $ null\left( {G^{\nu=1,2} } \right) $, and thus, an
increase in the values of $ RMS_{recon}^{\nu=1,2} $. Section 4.2.3
conclusively demonstrates the fact that the condition number of $
\tilde A^{\nu}  \ $ and the $ RMS_{recon}^{\nu} $ do not necessarily
increase simultaneously. Specifically, the case presented in Section
4.2.3 demonstrates a significant \textit{decrease} in the condition
numbers of $ \tilde A^{\nu=1,2}  \ $, that is accompanied by a
significant \textit{increase} in $ RMS_{recon}^{\nu=1,2} $.

This dichotomy may be explained by the fact that an
\textit{increase} in the value of $ RMS_{recon}^{\nu=1,2} $ is
always accompanied by a corresponding \textit{increase} in the RMS
error of encryption: $ RMS_{enc}^{\nu=1,2} $.  This trend is evident
in each of the case studies presented in this paper, and, has been
consistently observed in numerous simulation exercises.  As
indicated in the introduction of Section 4, the RMS error of
encryption ($ RMS_{enc}^{\nu} $) represents a more relevant metric
to assess the sensitivity of $ null\left( {G^\nu } \right) $ to
perturbations $ \delta G_{i,j}^\nu $, as compared to the condition
number of $ A^\nu $, within the framework of this statistical
encryption/decryption model.

This Section is concluded by providing a glimpse of the effects of
the instability of the eigenstructures of the ill-conditioned
operators. Figure 2 depicts the error of encryption: $ \left|
{\tilde p_c^\nu } \right\rangle $ - $ \left| {p^\nu }\right\rangle $
for the case study presented in Section 4.2.3, corresponding to $ \
k^{\nu=2} = -0.5 $ \. The highly oscillatory ("chaotic") profile in
Figure 2 is indicative of the extreme sensitivity of $ null\left(
{G^\nu } \right) $ to perturbations $ \delta G_{i,j}^\nu $ applied
to $ G^\nu $.

It may be noted that the code $ \left| {\tilde q^\nu } \right\rangle
$, projected into each host \textit{pdf}, comprises of a
\textit{ket} of dimension 396 created by a random number generator.
While one might expect the randomness inherent in $ \left| {\tilde
q^\nu } \right\rangle $ to be factored into the error of encryption,
further investigation concerning the oscillatory behavior is
required.  The \textit{ket} $ \left| {\tilde q^\nu } \right\rangle $
is subjected to a sorting operation, by employing the $
MATLAB^{\circledR} $ routine $ \ sort\left( \bullet \right) $.  This
results in a \textit{ket} $ \left| {\tilde q^\nu } \right\rangle $
of dimension 396, containing progressively increasing values defined
in $ \left[ {0,1} \right] $. The resulting error of encryption using
the sorted code is depicted in Figure 3.

As is observed, the error of encryption still displays a highly
pronounced oscillatory behavior, despite the randomness in the code
being mitigated.  Akin to the case depicted in Figure 2, this
oscillatory behavior is reflective of the instability induced in $
null\left( {G^\nu } \right) $ by perturbing 2 elements of $ G^\nu $
(having dimension $ 401 \times 401 $) by $ 3.0e-013 $ and $ 7.0e-013
$, respectively.
 Figure 4 depicts the total \textit{pdf} defined by (10),
corresponding to the case study presented in Section 4.2.3 for $
\nu=2 $.  As is expected, the total \textit{pdf} exhibits a highly
oscillatory ("chaotic") behavior. This is in stark contrast to the
smooth curves of the host \textit{pdf}'s depicted in Figure 1.

\begin{figure}[thpb]
    \centering
        \includegraphics[scale=0.7]{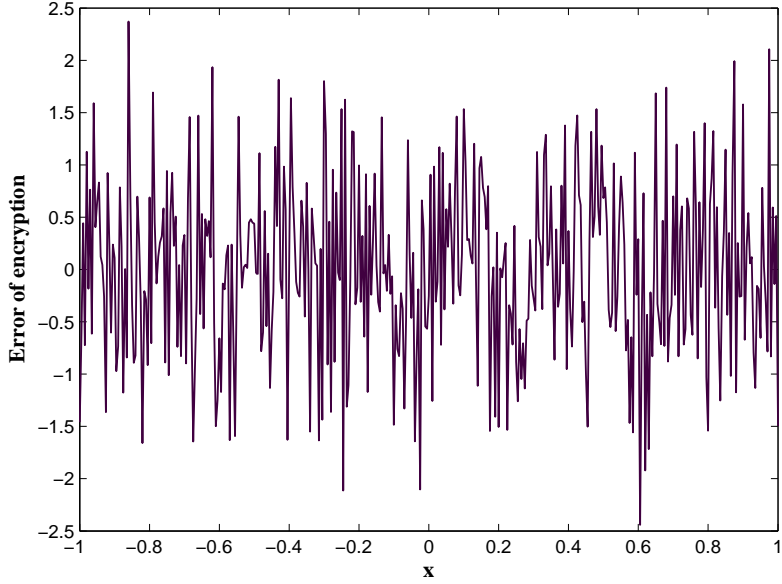}
    \caption{Plot of the error of encryption: $ \left| {\tilde p_c^\nu }
\right\rangle $- $ \left| {p_c^\nu } \right\rangle $.}
    \label{fig:Quadratic}
\end{figure}

\begin{figure}[thpb]
    \centering
        \includegraphics[scale=0.7]{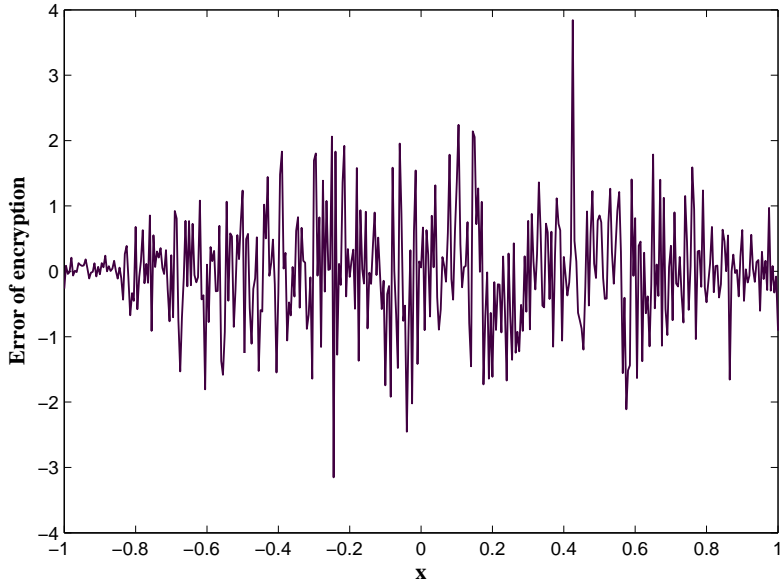}
    \caption{Plot of the error of encryption: $ \left| {\tilde p_c^\nu }
\right\rangle $- $ \left| {p_c^\nu } \right\rangle $ with sorted
code.}
    \label{fig:Quadratic}
\end{figure}

\begin{figure}[thpb]
    \centering
        \includegraphics[scale=0.7]{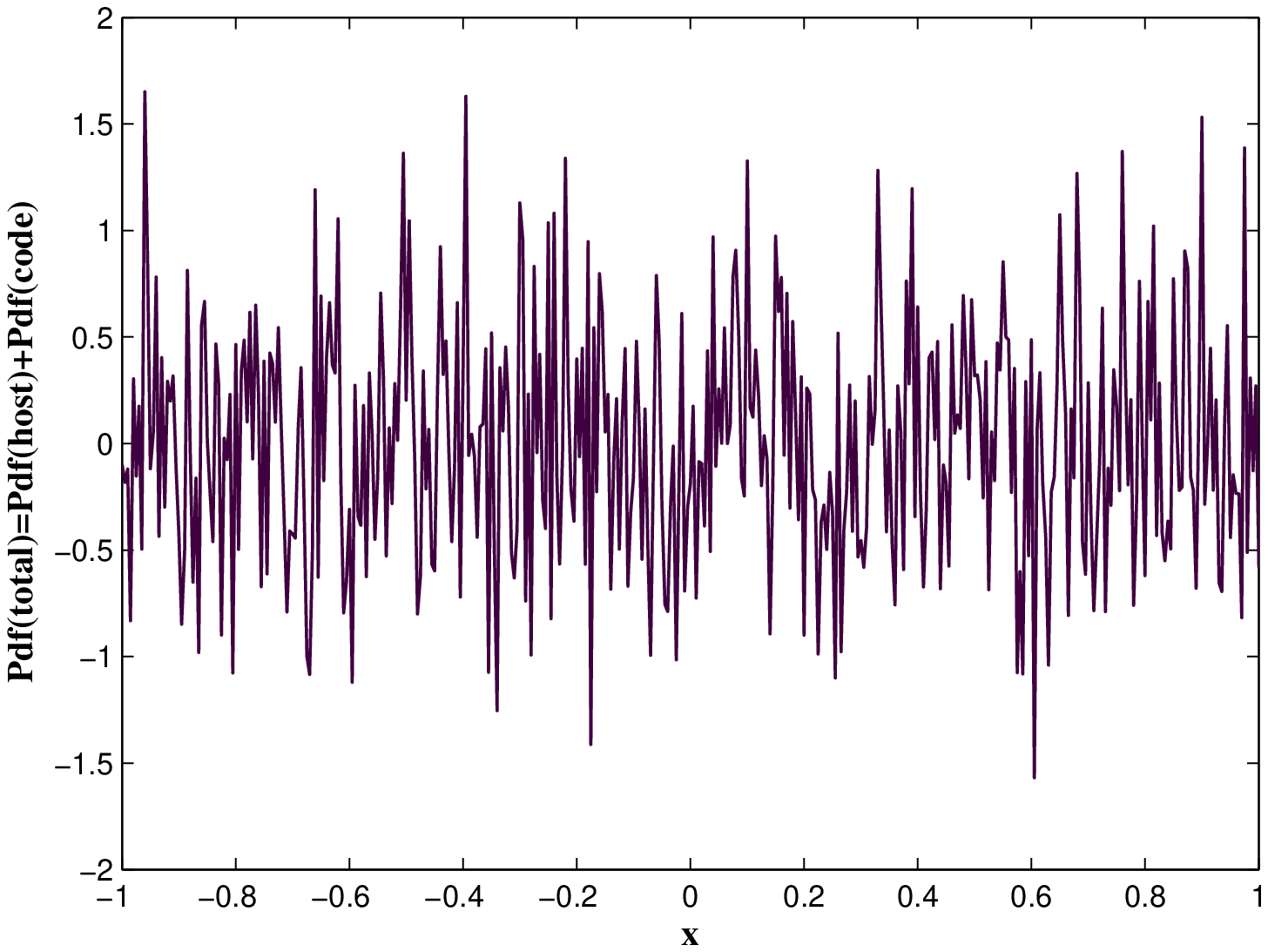}
    \caption{"Chaotic" nature of the total \textit{pdf} = host \textit{pdf} + code \textit{pdf}}
    \label{fig:Quadratic}
\end{figure}
\section{Summary and Conclusions}

A novel strategy to project covert information into a hierarchy of
statistical hosts has been presented.  This has been accomplished
within the ambit of the MaxEnt principle. The encryption/decryption
strategy relates the projection of the covert information to the
host \textit{pdf}'s.  This feature permits the statistical
encryption/decryption strategy to possess immense qualitative
flexibility, and, provide enhanced security to the covert information, as compared to a \textit{pdf} independent model
[8].

The \textit{pdf} dependent model, presented herein, sometimes
demonstrates an increased RMS error of reconstruction for decreased
values of the condition number of the constraint operator $ A^\nu $.
This seemingly counter-intuitive result is adequately explained with the aid
of the RMS error of encryption.  The RMS error of encryption is demonstrated to be a viable and
relevant metric to assess the sensitivity of $ null\left( {G^\nu }
\right) $ to perturbations $ \delta G_{i,j}^\nu $.

The statistical encryption/decryption strategy is platform
independent, and the process of recovery of the covert information
is accomplished with a very high degree of precision.  The small
amounts of data being transmitted through the covert channel as a
consequence of the transmission of the cryptographic keys $ \delta
G_{i,j}^\nu $ and the \textit{hierarchy indices} $ k^{\nu} $'s (see
Section 3.2), augers well for a coupling between the statistical
encryption/decryption model, and a quantum key distribution protocol
[13, 14].

A study extending the present work, by describing the statistical
encryption/decryption strategy within the framework of a
Fisher-Schr\"{o}dinger model, has been recently completed. Herein,
the Fisher information has been employed as the measure of
uncertainty. The host \textit{pdf}'s satisfy a time independent
Schr\"{o}dinger-like equation (TISLE) with an empirical
\textit{pseudo-potential}, that approximates a time independent
Schr\"{o}dinger equation (TISE) physical potential [7, 15] . The
TISLE inherits the \textit{energy states} of the TISE, within an
information theoretic context.  The encryption of covert information
is tantamount to projection of the code into different
\textit{energy states} of the TISLE. The \textit{hierarchy indices}
in the present paper are replaced by the TISLE \textit{energy
states}.

The Fisher-Schr\"{o}dinger model provides a quantum mechanical
connotation to the statistical encryption/decryption strategy.
Coupling the Fisher-Schr\"{o}dinger model with a quantum key
distribution protocol holds forth the prospect of achieving a
self-consistent \textit{hybrid} statistical/quantum mechanical
cryptosystem. This work will be shortly presented for publication.

Ongoing work is directed towards a two-pronged objective. First, the
projection of the covert information into $ null\left( {G^\nu }
\right) $ has been provided with an information theoretic basis.
Next, models to extend the work presented in this paper to the case
of steganography and information hiding have been developed.

By definition, steganography involves hiding data inside other
(host) data.  In steganography, the covert information is concealed
in a manner such that no one apart from the intended recipient knows
of the existence of the message.   This is in contrast to
cryptography, where the existence of the message is obvious, but the
meaning is obscured. The extension of the statistical
encryption/decryption model presented in this paper to steganography
involves encrypting image data into a \textit{cover image}.  In this
case, the average pixel intensities of the \textit{cover image}
constitute the incomplete constraints. These results will be
presented in future publications.

\textbf{Acknowledgements}

This work was supported by \textit{MSR contract} \textit{CSM-DI $ \
\& $ S-QIT-101107-2005}.  Gratitude is expressed to A. Plastino, L.
Rebollo-Neira, and B. R. Frieden for helpful discussions.


\end{document}